# Persistence of Monoclinic Crystal Structure in Three-Dimensional Second-Order Topological Insulator Candidate 1*T*'-MoTe$_2$ Thin Flake without Structural Phase transition

*Bo Su,# Yuan Huang,# Yan Hui Hou, Jiawei Li, Rong Yang, Yongchang Ma, Yang Yang, Guangyu Zhang, Xingjiang Zhou, Jianlin Luo, and Zhi-Guo Chen\**

B. Su, Prof. Y. Huang, Y. H. Hou, J. Li, Prof. R. Yang, Prof. Y. Yang, Prof. G. Zhang, Prof. X. Zhou, Prof. J. Luo, Prof. Z.-G. Chen
Beijing National Laboratory for Condensed Matter Physics
Institute of Physics, Chinese Academy of Sciences
Beijing 100190, China
Email: zgchen@iphy.ac.cn

B. Su, J. Li, Prof. G. Zhang, Prof. X. Zhou, Prof. J. Luo, Prof. Z.-G. Chen
School of Physical Sciences
University of Chinese Academy of Sciences
Beijing 100190, China

Y. H. Hou, Prof. Y. Ma
School of Materials Science and Engineering
Tianjin University of Technology
Tianjin 300384, China

Prof. Y. Huang, Prof. R. Yang, Prof. Y. Yang, Prof. G. Zhang, Prof. X. Zhou, Prof. J. Luo, Prof. Z.-G. Chen
Songshan Lake Materials Laboratory
Dongguan, Guangdong, 523808, China

Prof. G. Zhang, Prof. X. Zhou, Prof. J. Luo
Collaborative Innovation Center of Quantum Matter
Beijing, China

**A van der Waals material, MoTe$_2$ with a monoclinic 1$T$' crystal structure is a candidate for three-dimensional (3D) second-order topological insulators (SOTIs) hosting gapless hinge states and insulating surface states. However, due to the temperature-induced structural phase transition, the monoclinic 1$T$' structure of MoTe$_2$ would be transformed into the orthorhombic $T_d$ structure as the temperature is lowered, which hinders the experimental verification and the electronic applications of the predicted SOTI state at low temperatures. Here, we present systematic Raman spectroscopy studies of the exfoliated MoTe$_2$ thin flakes with variable thicknesses at different temperatures. As a spectroscopic signature of the orthorhombic $T_d$ structure of MoTe$_2$, the out-of-plane vibration mode $D$ at ~ 125 cm$^{-1}$ is always visible below a certain temperature in the multilayer flakes thicker than ~ 27.7 nm, but vanishes in the temperature range from 80 K to 320 K when the flake thickness becomes lower than ~ 19.5 nm. The absence of the out-of-plane vibration mode $D$ in the Raman spectra here demonstrates not only the disappearance of the monoclinic-to-orthorhombic phase transition but also the persistence of the monoclinic 1$T$' structure in the MoTe$_2$ thin flakes thinner than ~ 19.5 nm at low temperatures down to 80 K, which may be caused by the high enough density of the holes introduced during the gold-enhanced exfoliation process and exposure to air. The MoTe$_2$ thin flakes with the low-temperature monoclinic 1$T$' structure provide a material platform for realizing SOTI states in van der Waals materials at low temperatures, which paves the way for developing a new generation of electronic devices based on SOTIs.**

A topological insulator (TI), which is a type of quantum materials with time-reversal-symmetry-protected gapless states on the surfaces or along the edges and insulating states in the bulk, has attracted intensive attentions due both to its novel physical properties and potential applications in spintronic devices, terahertz detectors and quantum computing.[1-8] Lately, the concept of topological insulator was extended to a new class of systems called higher-order topological insulators (HOTIs): $d$-dimensional $n$th-order TIs have symmetry protected gapless ($d$-$n$)-dimensional boundary states (see **Figure** 1). Therein, intrinsic three-dimensional (3D) second-order topological insulators (SOTIs) exhibit topologically protected gapless states on the 1D hinges, but host insulating states on the 2D surfaces and in the 3D bulk.[9-17] Furthermore, it was suggested that the topologically protected 1D hinge state of 3D SOTIs sheds new light on the development of novel electronic applications based on Majorana bound states and surface quantum anomalous Hall effect, such as topological quantum computers and chiral circuit interconnects.[18,19] However, to date, few realistic materials have been identified experimentally as 3D SOTIs.[18,20] A natural question to ask is whether a broad class of 3D SOTIs can be found experimentally.

Very recently, a transition metal dichalcogenide $MoTe_2$ with the monoclinic crystal structure (i.e., $1T'$ phase with centrosymmetric space group $P2_1/m$) was theoretically predicted to be a candidate for 3D SOTIs.[21,22] However, when the bulk crystal of $1T'$-$MoTe_2$ is cooled below temperature $T \approx 250$ K, its monoclinic structure would transition into the orthorhombic structure (i.e., $T_d$ phase with non-centrosymmetric space group $Pmn2_1$) which was revealed to exhibit a type-II Weyl semimetal (WSM) ground state.[23-31] Therefore, accompanied with the temperature-induced structural phase transition from high-temperature $1T'$ to low-temperature $T_d$ phase, an electronic topological phase transition from a SOTI to a WSM is expected to take place in the bulk $MoTe_2$ crystal, i.e., the SOTI state in $1T'$-$MoTe_2$ would be destroyed at low temperatures. The occurrence of the structural (or topological) phase transition around $T \approx 250$ K in the bulk crystal of $1T'$-$MoTe_2$ means that (i) it is challenging to observe the predicted SOTI state in $1T'$-$MoTe_2$ because of the high-temperature-induced blurring of the energy dispersions of its gapless hinge states, and (ii) it is difficult to make use of the SOTI state in $1T'$-$MoTe_2$ to design a new generation of electronic devices working stably at low temperatures. Therefore, it is significant for the experimental identification and the electronic application of the SOTI state in $1T'$-$MoTe_2$ to maintain the monoclinic crystal structure at low temperatures. Up till now, the monoclinic $1T'$ crystal structure of $MoTe_2$ at low temperatures was mainly realized by applying ultrashort laser pulses, exerting mechanical pressure, or doping carriers.[32-34] It is worth noticing that the bulk crystal of $1T'$-$MoTe_2$ can be mechanically exfoliated into thin flakes because the bonding between $MoTe_2$ layers is van der Waals type. Nevertheless, whether the MoTe$_2$ thin flakes can still have the monoclinic 1$T'$ crystal structure at low temperatures remains elusive.

Raman spectroscopy is an experimental technique which is directly sensitive to the crystal structure.[35-42,44-53] Previous Raman spectroscopy studies of the bulk 1$T'$-MoTe$_2$ crystals indicate that in the high-temperature 1$T'$ phase at $T \geq 250$ K upon cooling (or at $T \geq 260$ K upon warming), the out-of-plane vibration mode $D$ around 125 cm$^{-1}$ is Raman in-active (i.e., only infrared active) and is absent in the Raman spectra due to the centrosymmetry of the monoclinic 1$T'$ structure, while in the low-temperature $T_d$ phase at $T < 250$ K upon cooling (or at $T < 260$ K upon warming), the out-of-plane vibration mode $D$ becomes both Raman- and infrared-active and can be probed by Raman spectroscopy owing to the centrosymmetry breaking in the orthorhombic $T_d$ structure (see the two vibration modes $D$ and $E$ in the Raman spectra of the MoTe$_2$ bulk crystal with the orthorhombic $T_d$ structure measured at $T = 80$ K in **Figure** 2a, the vibration mode $e$ in the Raman spectra of the MoTe$_2$ bulk crystal with the monoclinic 1$T'$ structure at $T = 300$ K in **Figure** 2b and the Raman spectra of the MoTe$_2$ bulk crystal in the energy range from 60 cm$^{-1}$ to 300 cm$^{-1}$ in Supplementary **Figure** 1 of Supporting Information).[35-40] Thus, the presence of the out-of-plane vibration mode $D$ can be regarded as a spectroscopic signature of the temperature-driven structural phase transition in MoTe$_2$ from the high-temperature monoclinic 1$T'$ structure to the low-temperature orthorhombic $T_d$ structure. Raman spectroscopy investigations of the thickness dependence and temperature evolution of the out-of-plane vibration mode $D$ enables us to gain insights into the fate of the monoclinic-to-orthorhombic phase transition in the MoTe$_2$ thin flakes at low temperatures.

In order to obtain the MoTe$_2$ thin flakes, we exfoliated the 1$T'$-MoTe$_2$ bulk crystals grown by vapor-transport method using iodine as the transport agent.[32] The MoTe$_2$ thin flakes were fabricated using the gold-enhanced exfoliation method.[41] **Figure** 2c displays three typical thicknesses of the exfoliated MoTe$_2$ flakes characterized by atomic force microscopy. Then, we used a 532 nm laser to measure the Raman spectra of the MoTe$_2$ thin flakes at different temperatures in the parallel-polarized configuration (i.e., the electrical field of the linearly polarized incident light is parallel to the electrical field of the linearly polarized scattered light, here). **Figure** 3 depicts the thickness evolution of the representative Raman spectra of the exfoliated MoTe$_2$ thin flakes measured in the temperature range from 80 K to 320 K (see the representative Raman spectra plotted in the energy range from 60 cm$^{-1}$ to 300 cm$^{-1}$ in Supplementary **Figure** 2 of Supporting Information and the Raman spectra of the MoTe$_2$ thin flakes with the thicknesses of ~ 31.1 nm and ~ 28.5 nm in Supplementary **Figure** 3a and 3b of Supporting Information). For the MoTe$_2$ flakes thicker than ~ 27.7 nm, the out-of-plane vibration mode $D$ around 125 cm$^{-1}$ is invisible in their Raman spectra at $T \geq 260$ K, but the mode $D$, together with the mode $E$ around 128 cm$^{-1}$, is present in their Raman spectra measured at $T < 260$ K (see **Figure** 3a-c), which indicates that the monoclinic-to-orthorhombic phase transition occurs in the MoTe$_2$ flakes thicker than ~ 27.7 nm. It is worth noticing that for the MoTe$_2$ thin flakes thinner than ~ 19.5 nm, a mode is present around 128 cm$^{-1}$ in the Raman spectra measured in the temperature range from 80 K to 320 K, while the mode $D$, which should appear around 125 cm$^{-1}$ in the Raman spectra of the MoTe$_2$ crystals with the orthorhombic $T_d$ structure, not only is invisible at $T \geq 260$ K, but also disappears at low temperatures down to 80 K (see **Figure** 3d-f), revealing that the vanishing of the monoclinic-to-orthorhombic phase transition and the persistence of the monoclinic 1$T'$ structure at temperatures above 80 K in the MoTe$_2$ thin flakes thinner than ~ 19.5 nm.

To check the thickness dependence of the monoclinic-to-orthorhombic transition in the MoTe$_2$ flakes, we plotted the temperature dependence of the representative Raman spectra of the exfoliated MoTe$_2$ thin flakes with the thickness varying from ~ 150 nm to ~ 13.8 nm in **Figure** 4 (see the Raman spectra of the MoTe$_2$ thin flakes measured at $T$ = 180 K, 160 K, 120 K and 100 K in Supplementary **Figure** 3c-f of Supporting Information). When 260 K $\leq T \leq$ 320 K, the mode $D$ is always invisible in the Raman spectra of the MoTe$_2$ flakes with the thickness ranging from ~ 150 nm to ~ 13.8 nm (see **Figure** 4a-c), which is in agreement with the present of the monoclinic 1$T'$ structure in the MoTe$_2$ bulk crystals at $T \geq 260$ K upon warming. By contrast, when 80 K ≤ $T$ < 260 K, the mode $D$ is visible in the Raman spectra of the MoTe$_2$ flakes thicker than ~ 27.7 nm, and then becomes absent in the Raman spectra of the MoTe$_2$ thin flakes thinner than ~ 19.5 nm (see **Figure** 4d-f), which also manifests the disappearance of the monoclinic-to-orthorhombic phase transition above 80 K and the existence of the low-temperature monoclinic 1$T$' structure in the MoTe$_2$ thin flakes thinner than ~ 19.5 nm.

To further confirm the existence of the low-temperature monoclinic crystal structure in the MoTe$_2$ thin flakes, we measured the Raman spectra of the MoTe$_2$ ultrathin flakes with the thicknesses lower than those shown in **Figure** 3 and **Figure** 4 at $T$ = 80 K. As displayed in **Figure** 5a, a mode is present around 128 cm$^{-1}$, but the mode $D$, which should be visible around 125 cm$^{-1}$ in the Raman spectra of the MoTe$_2$ crystals with the orthorhombic $T_d$ structure, is absent in the Raman spectra of the MoTe$_2$ ultrathin flakes with several representative thicknesses of ~ 10.9 nm, ~ 8.3 nm, ~ 5.7 nm, and ~ 3.3 nm, which indicates the persistence of the monoclinic 1$T$' structure in the MoTe$_2$ ultrathin flakes with the thicknesses down to ~ 3.3 nm at $T$ = 80 K. In **Figure** 5b, we plotted the false-color map of the intensities of the mode $D$ obtained by the Lorentzian fits as a function of thickness and temperature. Since the dark-blue color in **Figure** 5b represents the absence of the mode $D$ in the Raman spectra of the MoTe$_2$ thin flakes with the monoclinic 1$T$' structure, while the bright colors, such as yellow and green, show the presence of the mode $D$ in the Raman spectra of the MoTe$_2$ thin flakes with the orthorhombic $T_d$ structure, the false-color map of the intensity of the mode $D$ in **Figure** 5b can be regarded as the structural-phase diagram of the MoTe$_2$ flakes as a function of thickness and temperature.

A previously reported theoretical investigation of the structural phase transition in MoTe$_2$ indicates that (i) across the monoclinic-to-orthorhombic phase transition in the bulk MoTe$_2$ crystals, a net charge is transferred from the intralayer bonding state around the $Y$ point of the Brillouin zone to the interlayer antibonding states along the Γ-$A$ direction near Fermi energy, (ii) the net charge transfer lowers the total energy of the system, which can induce the occurrence of the monoclinic-to-orthorhombic phase transition in the bulk MoTe$_2$ crystals, (iii) the energy difference between the monoclinic phase and the orthorhombic phase is quite small (i.e., ~ 0.4 meV per unit cell for MoTe$_2$), (iv) the monoclinic (or orthorhombic) structure of the bulk MoTe$_2$ crystals can be stabilized by the hole (or electron) doping.[43] Therefore, the doping type and the doping level at room temperature are expected to induce different behaviors of the structural phase transition in MoTe$_2$, which can be supported by the observation of the different crystal structures of the thin flakes with the same thickness and the existence of the intermediate phase corresponding to neither the monoclinic phase nor the orthorhombic phase.[44-49] It is worth noticing that the phonon energies of the transition metal dichalcogenides can show significant dependence on the doping level and the doping type.[50] For 2$H$-MoTe$_2$, the out-of-plane vibration mode A$_{1g}$ exhibits a blue (or red) shift with the enhancement of the hole (or electron) concentration, while the A$_{1g}$ mode shows a red shift with the decrease in the flake thickness.[51] For the MoTe$_2$ flakes here, the out-of-plane vibration mode around 74 cm$^{-1}$ in **Figure** 5c displays a blue shift of ~ 2 cm$^{-1}$ as the flake thickness decreases from 21.1 nm to 3.3 nm, which implies that the hole concentration within the MoTe$_2$ flakes increases with the decrease in the flake thickness. Moreover, considering that (i) the MoTe$_2$ thin flakes in our manuscript were fabricated using the gold-enhanced exfoliation method (see ref. [41] or the Experimental Section in our manuscript), and (ii) the work function of gold is larger than that of MoTe$_2$,[52,53] electrons are expected to be transferred from our MoTe$_2$ thin flakes to the golden substrate in the process of exfoliating the crystals (see the schematic in **Figure** 5d), which can lead to the hole doping in our MoTe$_2$ thin flakes. In addition, an exposure of our MoTe$_2$ thin flakes to air can also result in the hole doping.[54-58] Therefore, we speculate that (i) when the MoTe$_2$ flakes, which are fabricated using the gold-enhanced exfoliation method and are exposed to air, become thin enough (i.e., the flake thickness here is lower than ~ 19.5 nm), the hole concentration within the MoTe$_2$ flakes is likely to be high enough to stabilize the monoclinic structure (i.e., the monoclinic-to-orthorhombic phase transition is absent in the MoTe$_2$ flakes thinner than ~ 19.5 nm at least above 80 K), and that (ii) when the MoTe$_2$ flakes fabricated by the gold-enhanced exfoliation method are thick enough (i.e., the flake thickness here is larger than ~ 27.7 nm), the hole concentration within the MoTe$_2$ flakes may be not high enough so that the monoclinic structure of the MoTe$_2$ flakes thicker than ~ 27.7 nm would transition into the orthorhombic structure (i.e., the monoclinic-to-orthorhombic phase transition can take place in the MoTe$_2$ flakes thicker than ~ 27.7 nm). In brief, the possible reason why the MoTe$_2$ flake thickness is special for the structural phase transition is that as the flake thickness decreases, the concentration of the holes introduced during the gold-enhanced exfoliation process and exposure to air is expected to become higher and higher so that the monoclinic structure of the MoTe$_2$ thin flake with the thickness below a critical value can be stable at low temperatures.

It is worth noticing that in the process of decreasing the MoTe$_2$ flake thickness by the gold-enhanced exfoliation method, the increase in the hole concentration (i.e., $n$) within the flakes may be accompanied with the enhancement of the local Coulomb repulsion $U$ since the average distance (i.e., $r$) between two charges becomes smaller. It was reported that the local Coulomb repulsion $U$ has an effect of stabilizing the orthorhombic crystal structure, while the monoclinic crystal structure can be stabilized by the hole doping (see Figure 3 of ref. [43]). Therefore, the competition between the hole doping and the local Coulomb repulsion should play a significant role in determining whether the MoTe$_2$ flake goes into the monoclinic structure or the orthorhombic structure. Since (i) the local Coulomb repulsion $U$ can be regarded to be approximately proportional to the inverse (i.e., $1/r$) of the average distance between two charges, and (ii) the hole concentration (i.e., $n$) of the MoTe$_2$ flake is approximately linear with the inverse ($1/V$) of the flake volume ($V \propto r^3$) and is also proportional to the inverse ($1/r^3$) of the $r$ cubed, the local Coulomb repulsion $U$ can be roughly deemed to be proportional to the hole concentration to the one-third power, i.e., $U \propto n^{1/3}$. The approximately linear relationship between $U$ and $n^{1/3}$ means that (i) when the concentration (i.e., $n-n_0$) of the doped holes increases (due to a decrease in the flake thickness) but is not high enough (i.e., $(n-n_0)/n_0 \ll 1$, here $n_0$ is the hole concentration before doping), the local Coulomb repulsion $U$ would increase so sharply that the effect of the local Coulomb repulsion $U$ on the crystal structure can be comparable to that of the hole doping, which may destroy the hole-doping-induced stabilization of the monoclinic crystal structure and thus may result in the weak temperature dependence of the structural transition temperature of the MoTe$_2$ flakes thicker than 27.7 nm; and (ii) when the concentration of the doped holes continues to increase (owing to the continuous decrease in the flake thickness), the growth velocity of the local Coulomb repulsion $U$ would decrease notably and then would be lower than the constant growth velocity of the concentration of the doped holes, which may ultimately make the effect of the hole doping on the crystal structure have an advantage over that of the local Coulomb repulsion $U$ and therefore may lead to the abrupt lowering of the structural-phase-transition temperature in the MoTe$_2$ flakes with the thicknesses ranging from 19.5 nm to 27.7 nm.

In summary, using Raman spectroscopy, we have investigated the crystal structures of the exfoliated 1$T$'-MoTe$_2$ thin flakes with different thicknesses as a function of temperature. When the MoTe$_2$ flake thicknesses are larger than ~ 27.7 nm, the out-of-plane vibration mode $D$ appears around 125 cm$^{-1}$ in the Raman spectra at $T$ < 260 K and then becomes invisible at $T$ ≥ 260 K, which indicates that the MoTe$_2$ flakes thicker than ~ 27.7 nm undergo the structural phase transition from the monoclinic 1$T$' structure to the orthorhombic $T_d$ structure. When the MoTe$_2$ flake thicknesses are thinner than ~ 19.5 nm, the mode $D$ is always absent in the Raman spectra even though the temperature is increased from 80 K to 320 K, which demonstrates the absence of a temperature-induced structural phase transition and the retention of the monoclinic 1$T$' structure at low temperatures down to 80 K in the MoTe$_2$ flakes thinner than ~ 19.5 nm. The possible reason for the persistence of the monoclinic 1$T$' structure at low temperatures is that as the MoTe$_2$ flake thickness is decreased below a critical value, the concentration of the holes introduced during the gold-enhanced exfoliation process and exposure to air is expected to become high enough to stabilize the monoclinic structure. Our work not only constitutes a significant step towards the experimental realization of the predicted SOTI state in the van der Waals material 1$T$'-MoTe$_2$ but also paves the way for the quest of SOTI states in realistic materials at low temperatures.

# Experimental Section

*Synthesis and Exfoliation of the 1T'-MoTe$_2$ bulk crystals*: The 1T'-MoTe$_2$ single crystals were prepared by the chemical vapor transport method.[32] Firstly, the raw materials of the polycrystalline MoTe$_2$ were prepared. The Mo and Te powders with a chemical stoichiometric ratio of 1:2 were firstly mixed and pressed into pellets. The pellets were then sealed in an evacuated quartz tube and heated to 800 °C within 20 hours. After annealing for 7 days, the tube was quenched into ice water quickly. At last, the prepared precursor and transport agent I$_2$ were sealed in an evacuated quartz tube and placed in a two-zone tube furnace with a temperature gradient from 1000 °C to 900 °C, and kept that temperature gradient or 7 days. At the end of the sequence, this quartz tube was quenched in the ice water. The 1T'-MoTe$_2$ thin flakes were fabricated using the gold-enhanced exfoliation method.[41] Firstly, a thin adhesion metal layer of Ti was evaporated on the Si/SiO$_2$ substrate using an electron evaporation system (Peva-600E), then a thin Au layer was deposited onto the pre-prepared metal layer of Ti. After that, the 1T'-MoTe$_2$ single crystals were mechanically exfoliated using the white tape (3 M scotch) and were cleaved for several times along the c-axis. Lastly, the freshly cleaved layered 1T'-MoTe$_2$ crystals together with the 3 M scotch tape were put onto the processed substrate and pressed vertically using a gentle pressure for about 1 minute, then the tape was peeled off from the substrate.

*Polarized Raman measurements:* The Raman spectroscopy measurements were carried out on a HORIBA LabRAM HR Evolution Raman spectrometer using a 532 nm laser. The spectroscopy is acquired in a backscattering geometry with the configurations of the incident and scattered photons polarized parallel to each other. For the Raman measurements in the temperature range from 80 K to 320 K, the sample temperature was controlled using a nitrogen cooled Linkam TS600 hot stage.

# Acknowledgements

#B.S. and Y.H. contributed equally to this work. Z.-G.C. conceived and supervised this project. B.S. and Y.H.H. carried out the Raman experiments. Y.H. exfoliated the bulk crystals

into the thin flakes. J.L., R.Y., and B.S. did the AFM measurements. Z.-G.C., B.S., Y.Y., Y.M., G.Z., X.Z., and J.L. analyzed the data. Z.-G.C. and B.S. wrote the paper. We thank Zhijun Wang, Young-Woo Son, Hyun-Jung Kim and Sang-Hoon Lee for very helpful discussions. The authors acknowledge support from the National Key Research and Development Program of China (Projects Nos. 2017YFA0304700, 2016YFA0300600, 2019YFA0308000, 2018YFA0704201, and 2018YFB0703500), the National Natural Science Foundation of China (Projects Nos. 12022412, 11874405, 62022089, and 11704401), the strategic Priority Research Program of Chinese Academy of Sciences (Project No. XDB33000000), and the Pioneer Hundred Talents Program of the Chinese Academy of Sciences.

## Conflict of Interest

The authors declare no conflict of interest.

**Figures, Figure Legends**

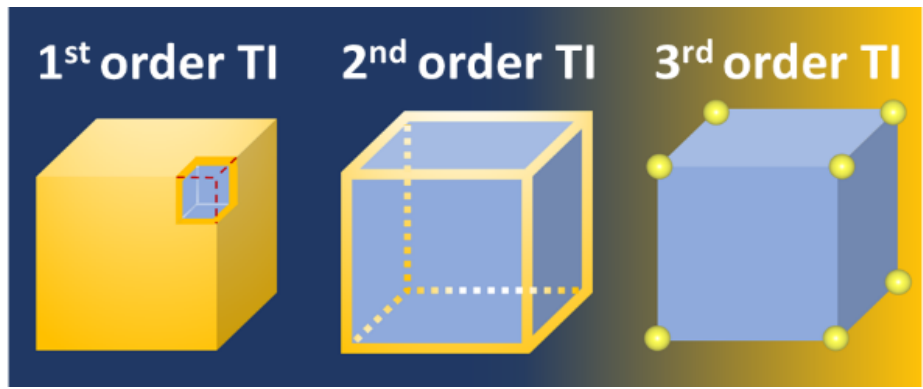

**Figure 1.** Schematic of the topologically protected states in three-dimensional (3D) topological insulators (TIs). The first-, second- and third-order 3D TIs have the topologically protected states (shown in bright yellow) on their 2D surfaces (left panel), 1D hinges (middle panel) and 0D corners (right panel), respectively.

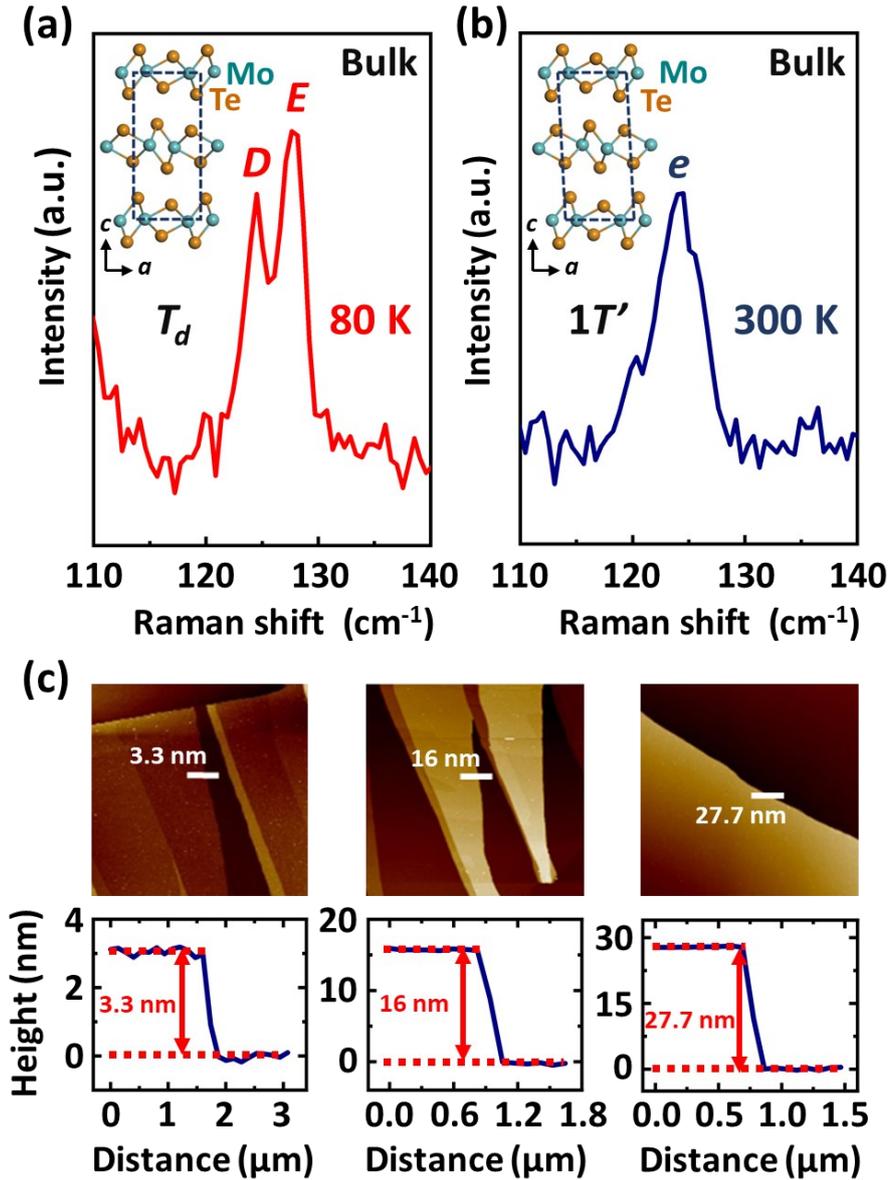

**Figure 2.** a) The two vibration modes *D* and *E* in the Raman spectrum of the MoTe$_2$ bulk crystal with the orthorhombic $T_d$ structure measured at *T* = 80 K. b) The vibration mode *e* in the Raman spectrum of the MoTe$_2$ bulk crystal with the monoclinic 1*T'* structure at *T* = 300 K. The up left insets in (a) and (b) show the $T_d$ and 1*T'* structure of MoTe$_2$, respectively. c) Upper panels: atomic-force-microscopy images of the MoTe$_2$ flakes. Lower panels: flake thicknesses along the white lines on the upper panels.

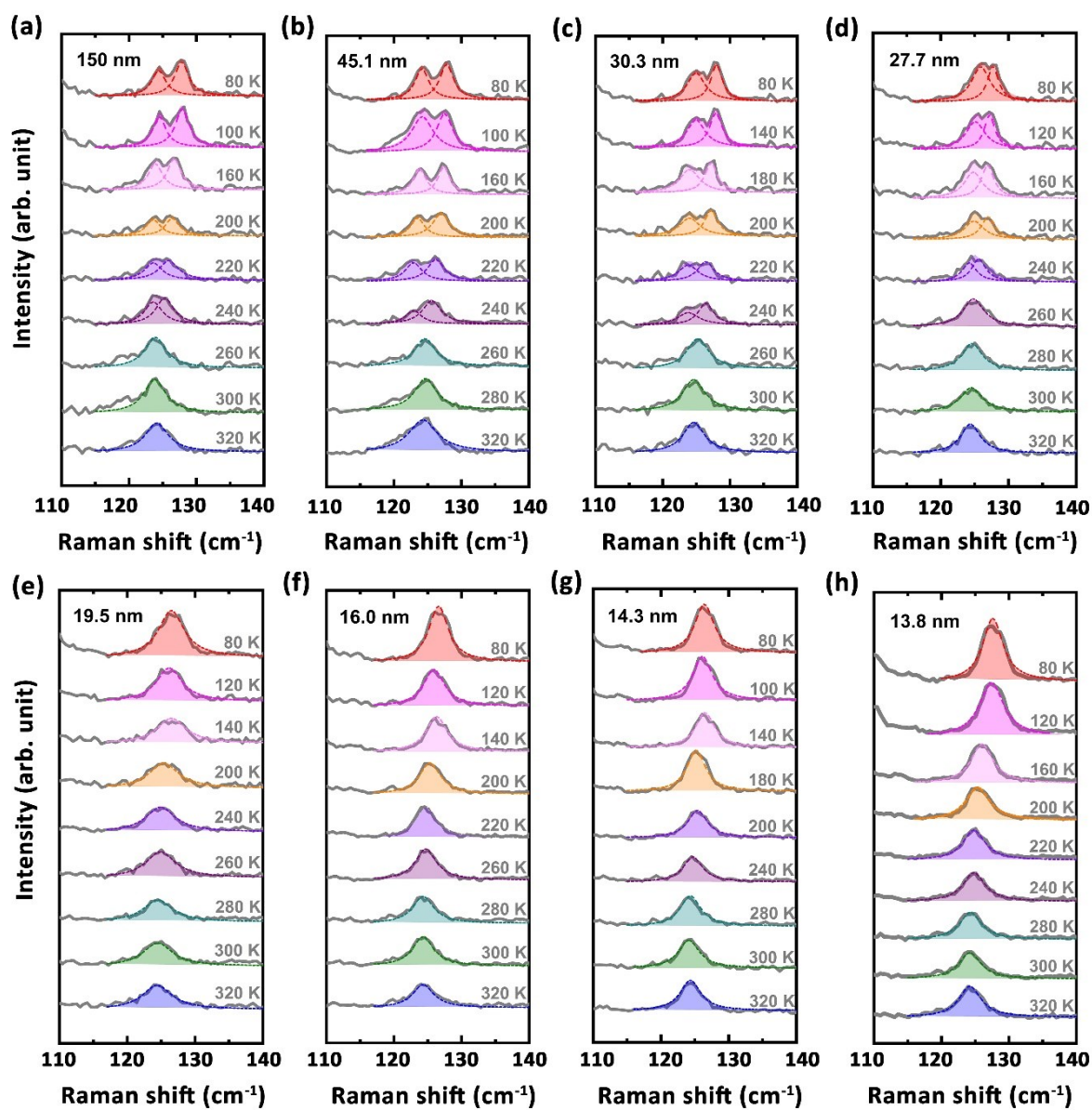

**Figure 3.** Thickness evolution of the representative Raman spectra of the MoTe$_2$ thin flakes measured in the temperature range from 80 K to 320 K. The dashed colored curves show the Lorentzian fits to the peak-like features.

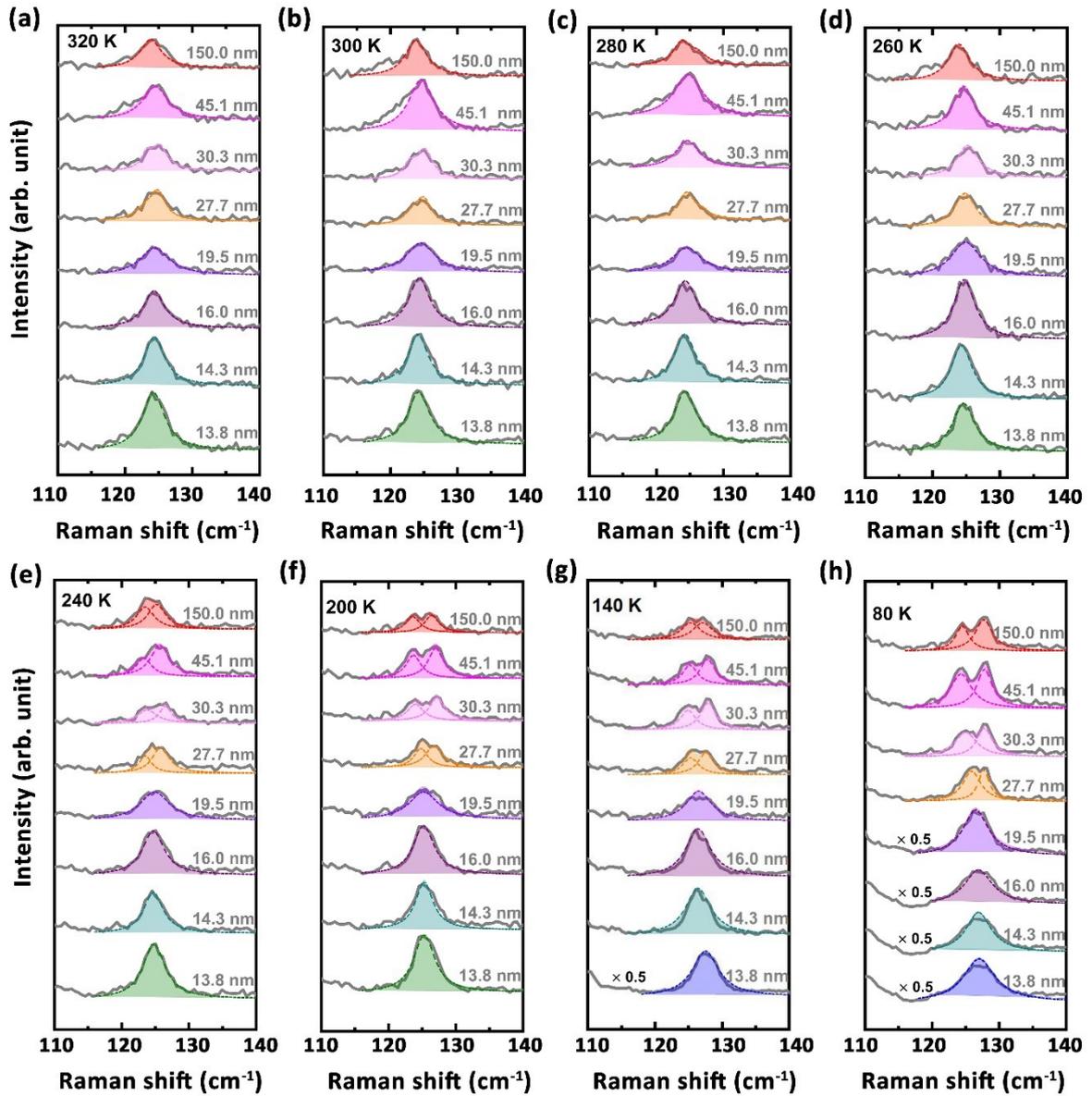

**Figure 4.** Temperature dependence of the representative Raman spectra of the MoTe$_2$ thin flakes with the thickness varying from ~ 150.0 nm to ~ 13.8 nm. The dashed colored curves show the Lorentzian fits to the peak-like features. The Raman spectra labelled with "× 0.5" in (e) and (f) are shown with the half of the intensities.

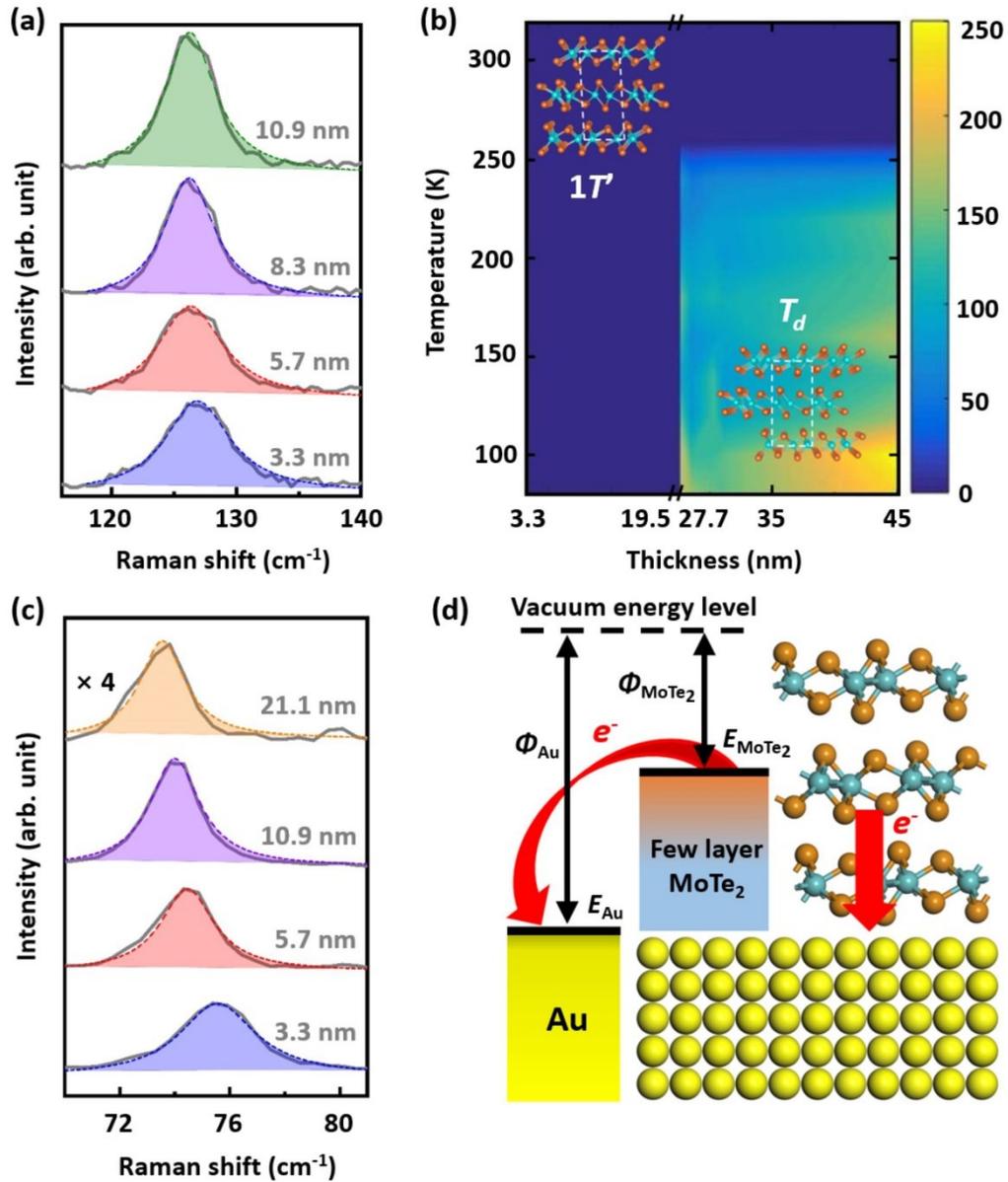

**Figure 5.** a) Raman spectra of the MoTe$_2$ ultrathin flakes with the thicknesses of ~ 3.3 nm, ~ 5.7 nm, ~ 8.3 nm and ~ 10.9 nm measured at $T$ = 80 K. The dashed curves show the Lorentzian fits to the peaks. b) Color scale map of the mode $D$ intensities obtained by the Lorentzian fits as a function of thickness and temperature. c) Out-of-plane vibration mode around 74 cm$^{-1}$ in the Raman spectra of the MoTe$_2$ ultrathin flakes with the thicknesses of ~ 3.3 nm, ~ 5.7 nm, ~ 10.9 nm and ~ 21.1 nm measured at $T$ = 80 K. The dashed curves show the Lorentzian fits to the peaks. The vibration mode labelled with "× 4" in (c) is displayed with the fourfold intensity. d) Schematic of the transfer of electrons from the MoTe$_2$ thin flake to the golden substrate in the exfoliating process. Here, $\phi_{\text{MoTe2}}$ and $\phi_{\text{Au}}$ represent the work functions of MoTe$_2$ and gold, respectively.